\documentclass[aps,pre,twocolumn]{revtex4}

\usepackage{graphicx}
\usepackage{amsmath,amscd,amssymb}

\begin{document}

\title{The vapor-liquid interface potential of (multi)polar fluids\\ and its influence on ion solvation}

\author{Lorand  Horv\'{a}th}
\affiliation{Universit\'{e} de Toulouse; UPS; Laboratoire de Physique Th\'{e}orique (IRSAMC); F-31062 Toulouse, France and CNRS; LPT (IRSAMC); F-31062 Toulouse, France}
\affiliation{Faculty of Physics, University ``Babe\c{s}-Bolyai'', 400084 Cluj-Napoca, Romania}
\author{Titus  Beu}
\affiliation{Faculty of Physics, University ``Babe\c{s}-Bolyai'', 400084 Cluj-Napoca, Romania}
\author{Manoel  Manghi}
\affiliation{Universit\'{e} de Toulouse; UPS; Laboratoire de Physique Th\'{e}orique (IRSAMC);
F-31062 Toulouse, France and CNRS; LPT (IRSAMC); F-31062 Toulouse, France}
\author{John Palmeri}
\affiliation{Universit\'{e} de Toulouse; UPS; Laboratoire de Physique Th\'{e}orique (IRSAMC);
F-31062 Toulouse, France and CNRS; LPT (IRSAMC); F-31062 Toulouse, France}
\affiliation{{Laboratoire Charles Coulomb UMR 5221 CNRS-UM2,
D\'{e}partement Physique Th\'{e}orique, Universit\'{e} Montpellier 2, Place Eug\`{e}ne Bataillon - CC070 F-34095 Montpellier Cedex 5 France}}
\altaffiliation[Current address]{}
\email{john.palmeri@univ-montp2.fr}

\date{\today}

\begin{abstract}

The interface between the vapor and liquid phase of
quadrupolar-dipolar fluids is the seat of an electric interfacial potential
whose influence on ion solvation and distribution is not yet fully
understood. To obtain further microscopic insight into water specificity we first present extensive classical molecular dynamics simulations of a series of model liquids with
variable molecular quadrupole moments that interpolates between SPC/E water
and a purely dipolar liquid. We then pinpoint the essential role played by the
competing multipolar contributions to the \textit{vapor-liquid} and the \textit{solute-liquid} interface potentials in determining an important ion-specific \textit{direct} electrostatic contribution to the ionic solvation
free energy for SPC/E water{---dominated by the quadrupolar and dipolar contributions---}beyond the dominant \textit{polarization} one.
{ Our results show that the influence of the \textit{vapor-liquid} interfacial potential on ion solvation is strongly reduced  due to the strong partial cancellation brought about  by the competing \textit{solute-liquid} interface potential.}

\end{abstract}
\maketitle
\section{Introduction}
\label{intro}
At the macroscopic interface between a
liquid ($l$) and its vapor ($v$) phase there is a spatial inhomogeneity that
induces a charge imbalance, producing an electric field and consequently a
potential difference across the interface, $\phi_{lv} =\phi_l-\phi_v$. Despite extensive molecular simulation studies at both the
classical and quantum mechanical levels over the past few decades~\cite{1,2,3,4,5,6,7,8,9,10,11,12,13,14}, a
complete understanding of this potential, how it depends on the
characteristics of the fluid studied, and its role in the solvation of ions
is not yet at hand~\cite{14a,14b}. Such an understanding has become a pressing matter,
because there is currently much interest in constructing mesoscopic models
of electrolytes near vapor-liquid interfaces and solid membrane surfaces and in nanopores~\cite{14a,14b,14c,14d,15,16,17,18,18a,19,20}.
It is also now clear that the value of the
interface potential observed experimentally depends on the probe used:
although electron diffraction and holography techniques may measure the full
interface potential, electrochemical techniques involved in ion solvation
seemingly do not~\cite{12,13,14,14b}.

Until now mesoscopic approaches to ion distribution have either completely neglected the contribution of the interfacial potential~\cite{15,18,18a,19}  or,  { as already  demonstrated in \cite{17} and discussed in detail here}, severely overestimated its importance by treating  finite size solutes as point test charges~\cite{16}.
For finite size ions a second microscopic solute-liquid interface potential, $\phi_{ls} =\phi_l -\phi_s$, exists, defined as the potential difference between the bulk liquid ($l$) and the center of a  neutral
solute ($s$) (fig.~\ref{fig1})  [whose size is determined in classical Molecular Dynamics (MD) simulations by the short range repulsion of the Lennard-Jones potential].
{   This second solute-liquid contribution is missed if the ions are approximated as point test charges.}
The potentially important role played by this microscopic potential in determining ion distribution near inhomogeneities needs to be clarified~\cite{10,10a} in order to provide deeper theoretical insight into  both molecular simulations and experimental results. Furthermore,  the coupling between the quadrupolar and dipolar contributions to the interface potential  and their respective roles in governing ion distribution need to be reconsidered.   To do so  we present extensive classical molecular dynamics simulations of model liquids  that interpolate between SPC/E water (a classical three site partial charge model ~\cite{spce}) and a purely dipolar liquid.

In physical terms our study can be viewed
as part of the quest, still far from complete, for the physical components of the
position dependent ionic Potential of Mean Force (PMF), $\Phi (\mathbf
r)$, near dielectric interfaces and surfaces arising from solvent-ion and ion-ion
interactions after the solvent degrees of freedom have been integrated out~\cite{14a,15,16,17,18,18a,19}.
We focus  uniquely on the poorly understood role played by the interfacial potential in determining the electrostatic contribution to ion solvation.
{   Indeed, the extremely large discrepancy between the dilute limit ionic PMF  obtained from MD simulations and those predicted using an approximate mesoscopic approach incorporating the contribution of the interfacial potential in the point ion approximation led the authors of \cite{17} to completely abandon their dilute limit mesoscopic approach; they opted rather for extracting the dilute limit ionic PMF directly from MD simulations and then injecting it into a generalized Poisson-Boltzmann equation to study salt concentration effects. In more recent work concerning the optimization of the MD parameters of a non-polarizable model by fitting to experiment, the same authors attempted to get around the ambiguities  plaguing  the contribution of the vapor-liquid  interfacial potential in determining the electrostatic contribution to ion solvation by fitting only quantities independent of this contribution \cite{19a}.}
Although  other contributions to the ionic PMF and solvation-free
energy, such as the hydrophobic and dispersion ones,  may play non-negligible roles and therefore be important for
interpreting molecular simulations and understanding experiments,
these contributions will not be considered here (as they are already  fairly well understood thanks to recent progress in this area)~\cite{15,16,17,18,18a,19,19a}.

One major impediment to
obtaining the physically identifiable mesoscopic contributions to the ionic
PMF, $\Phi $, arises from questions concerning the amplitude and sign of the
vapor-liquid interface potential and the role it plays in  determining the ionic solvation free energy.  In order to address these questions we  compare a direct evaluation from MD simulations of the two relevant electrostatic contributions to the ionic solvation free energy for SPC/E water---a \textit{direct interfacial} one that does not account for solvent polarization  due to the ionic charge and a \textit{polarization} one that does---with simplified approaches previously adopted in the literature  (namely,  a direct one approximating ions as point test charges and a simple Born-type polarization approximation,   defined below).

\section{ Vapor-Liquid interface potential and ionic PMF:  state-of-the-art}
\label{vl}

\begin{figure}
\resizebox{\columnwidth}{!}{\includegraphics{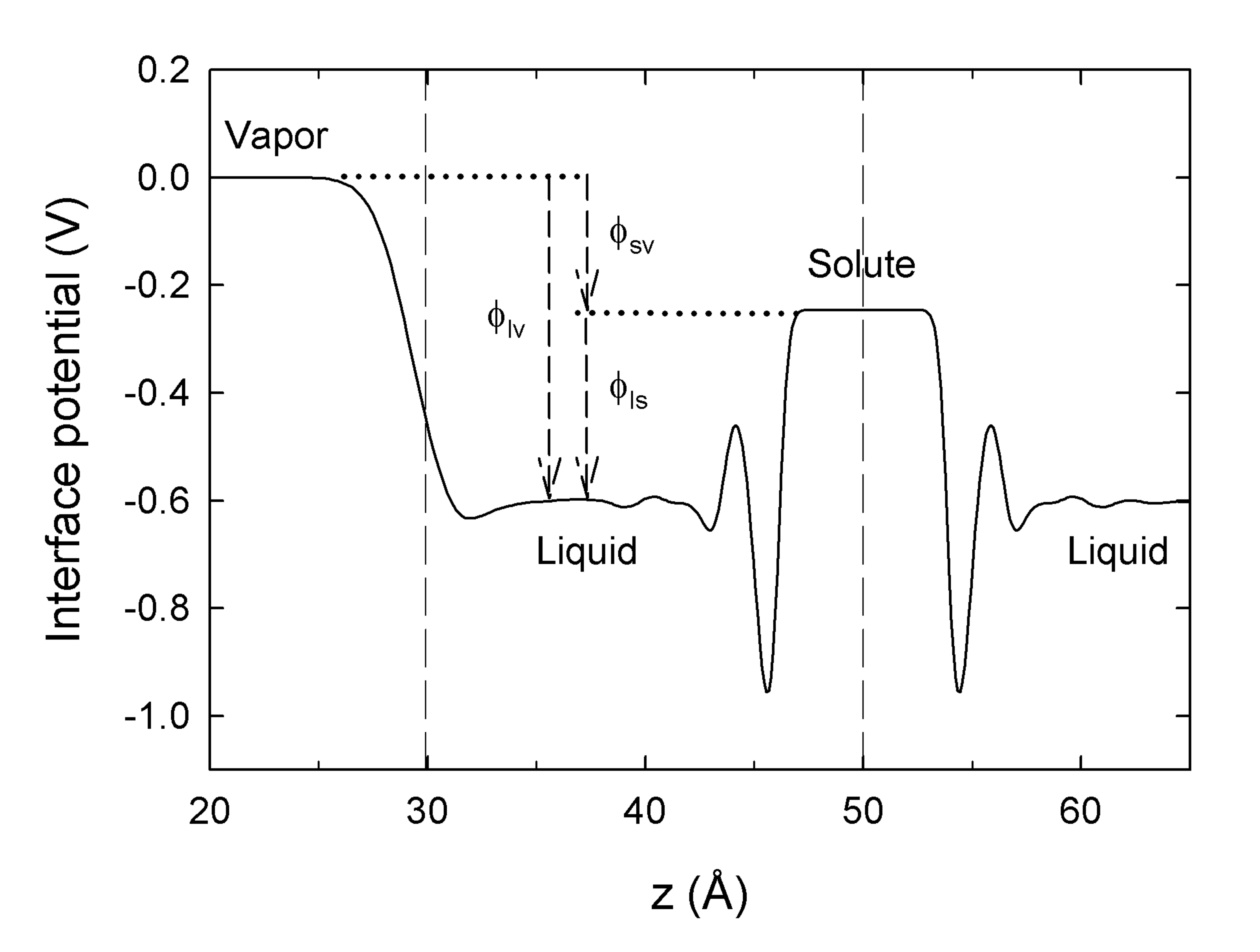}}
\caption{Electric potential variation, $\delta \phi (z)$, across the vapor-liquid and the liquid-solute interfaces for the neutral LJ
solute I$^{0}$ immersed in liquid water (SPC/E), at $z=50$~\AA. The Gibbs
dividing surface (GDS) of the macroscopic interface and the solute position
are indicated by dashed vertical lines.}
\label{fig1}
    \end{figure}

Near a planar vapor-liquid  interface  the local ion concentration can be expressed in terms of the PMF, $\Phi (z)$, as  $\rho_i (z)=\rho_{il} \exp [-\Phi (z)/k_{\rm B}T]$, where $z$ is the normal coordinate and $\rho _{il}$ is the ionic
concentration in the bulk liquid (where   $\Phi $ is taken to vanish). The
total ion solvation free energy can then be expressed as $\Delta
G^{\rm ion}=-\Phi (z_v )$, where $z_v$ is in the vapor phase (see fig.~\ref{fig1}). In theoretical studies of both
vapor-liquid water interfaces and membrane-liquid surfaces, it has sometimes
been hypothesized~\cite{16,17,18,18a} that the bare interface potential enter the PMF
via a simple \textit{direct} electrostatic contribution ${\Phi }'_{\rm pot}(z)=q [\delta \phi (z)-\phi _{lv}]$, where $q$ is the ion charge, $\delta
\phi (z)=\phi (z)-\phi_v$ is the local value of the potential
difference and $\phi _{lv} =\delta \phi (z_l )$ ($z_l$ is
at the center of the liquid slab far from the interface). This approach, which amounts to treating a finite size ion as a point test charge $q$~\cite{16,17,18,18a},  is critically examined here for SPC/E water.

Classical molecular dynamics (MD) simulations predict potentials on the
order of $-0.5$~V for both vapor-liquid interfaces and membrane-liquid
surfaces and, if used in the point ion approximation, $\Phi'_{\rm pot}$, would seemingly yield the
dominant contribution ($\sim 20~k_{\rm B} T$ for monovalent ions) to the
PMF over a substantial part of the interfacial region~\cite{17}. This approximation,
however,   predicts  incorrect results for the PMF and corresponding ion density, when compared with MD simulations, both in the infinitely dilute limit {   (as already pointed out in \cite{17})} and when incorporated into a modified Poisson-Boltzmann approach (to study higher electrolyte concentrations~\cite{16}): neither the strong build-up of anions near a strongly hydrophobic uncharged
surface~(\cite{16}, Fig.~4a), nor the variations in the dilute limit of the PMF near a membrane
surface~(\cite{17}, Fig.~3) predicted by this approach are in agreement with MD
simulations. This approximation also  yields a very substantial, albeit  seemingly undetected,  direct contribution to the ionic free energy of solvation,
    \begin{equation}
\label{eq0}
\Delta G'_{0}= -\Phi'_{\rm pot}(z_v) = q \phi_{lv}.
    \end{equation}
on the order of $ 25~k_{\rm B} T$.
Disturbingly, the reasonable agreement between the
experimental results for the surface tension of electrolyte solutions and certain promising mesoscopic theoretical approaches that
neglect the interface potential completely would be severely disrupted if such large
interfacial potentials were taken into account~\cite{18,18a}. This situation becomes
even more complicated if one considers that more ``realistic'' quantum
mechanical calculations can lead to \textit{positive} interface potentials of much higher
amplitude (+3~eV)~\cite{11,12,13}, but no signature of such
a potential is seen in recent \textit{ab initio} simulations of ion solvation~\cite{14}.

\section{Ionic free-energy of solvation}
The  electrostatic
(ES) contributions to the total ion solvation free energy for the models
studied here can be extracted directly from the MD simulations via
    \begin{equation}
\label{eq1}
\Delta G_{\rm ES}^{\rm ion} =\Delta G_0 +\Delta G_{\rm pol}
    \end{equation}
with
    \begin{equation}
\label{eq1a}
\Delta G_0 = q\phi_{sv} = q(\phi_{lv} - \phi_{ls})
    \end{equation}
and
    \begin{equation}
\label{eq1b}
\Delta G_{\rm pol} \simeq q (\phi_{sv}^{\rm ion} -\phi_{sv})/2
    \end{equation}
where  $\phi _{sv} =\phi _s -\phi _v =\phi_{lv} -\phi _{ls}$ and $\phi _{sv}^{\rm ion} =\phi _{lv} -\phi_{ls}^{\rm ion}$ are, respectively, the total vapor-liquid-solute potential variations for
neutral and charged solutes. The ion solute-liquid interface
potential, $\phi _{ls}^{\rm ion} =\phi _l^{\rm ion} -\phi _s^{\rm ion}$ is the potential difference between the bulk liquid and the center of the charged ion (with the bare Coulomb potential due to the ion itself
subtracted out). The first, \textit{direct}, term, $\Delta G_0 =q\phi _{sv}$, may be
regarded as the electrostatic free energy of solvation of an ion placed in
the potential $\phi _{sv}$ created around its corresponding \textit{neutral}
counterpart; the second \textit{polarization} term, $\Delta G_{\rm pol} $, obtained from an
approximate generalized ``charging'' method~\cite{22,23}, arises from the
response of the solvent to  the ion (which generates an overall
potential variation $\phi _{sv}^{{\rm ion}} $ much larger in amplitude
than $\phi _{sv} )$~\cite{23}. Because $\phi _{lv}$ is ion independent, the polarization contribution can be simplified to $\Delta G_{{\rm pol}}
\simeq q(\phi _{ls} -\phi _{ls}^{\rm ion})/2$ and therefore be obtained from bulk simulations (this  contribution is the microscopic analog of the mesoscopic Born one presented below).
{     Since $(\phi _{ls} -\phi _{ls}^{\rm ion}) \propto q$ vanishes in the limit $q\rightarrow 0$, $\Delta G_{{\rm pol}} \propto q^2$, as required for a polarization contribution (and also seen in the usual mesoscopic Born term below).}

\subsection{Mesoscopic Born model}
Within the mesoscopic Born model, an ion is modeled as a point charge $q$
sitting in its spherical cavity of effective radius $R_i $ in bulk water
treated as a continuum of dielectric constant $\varepsilon _w $. The radial
electric potential around the central ionic charge, $\varphi_{\rm ion}
(r)$, determines the Born approximation for the polarization contribution
via
    \begin{equation}
\label{eq2}
\Delta G_{\rm pol}^{\rm B} =\frac{q}{2}\mathop {\lim }\limits_{r\to 0}
[\varphi _{\rm ion} (r)-\varphi _0 (r)]=\frac{q^2}{8\pi
\varepsilon _0 R_i }\left(\frac1{\varepsilon _w} - 1 \right)
    \end{equation}
where $\varepsilon _0$  is the vacuum permittivity, $\varphi _0 (r)$ is the bare Coulomb potential, and $\varepsilon _w \simeq 78$ is the dielectric constant of bulk water at room temperature. Although this type of
polarization contribution ($\sim 100-180~k_{\rm B} T$) typically dominates the ionic
solvation free energy for the ions studied here, it is neither clear how
accurate the simple Born approximation is (due to the neglect of potentially important
ion-solvent correlations near the ion), nor how to choose best $R_i $.

Furthermore, despite its dominant role in the global ionic free energy of
solvation, the polarization contribution to the local ionic PMF, $\Phi(z)$, is seemingly not the dominant
contribution over a significant part of the
interfacial region, which means that the role of other contributions must be
clarified~\cite{17}. Although a Born-type polarization term is commonly
incorporated in mesoscopic approaches to the free energy of solvation (or
PMF), the direct term, arising from the bare interfacial potential, is
either completely neglected without justification~\cite{15,18,18a,19} or strongly
overestimated by incorrectly assuming  $ \phi _{ls} = 0$ in $\Delta G_0$ (eq.~\ref{eq1a}), which leads to the  approximation $\Delta G'_0 =q\;\phi _{lv}$ (eq.~\ref{eq0}) for the direct contribution
[or $\Phi'_{\rm pot} (z)$ in the PMF, which includes only the
contribution of the vapor-liquid interface \cite{17,16} and neglects entirely that of the solute-liquid one] (see fig.~1).

\section{Models and methods}
We compare a
direct evaluation from MD simulations of the two terms contributing
to $\Delta G_{\rm ES}^{\rm ion} $, namely $\Delta G_0 $ and $\Delta
G_{\rm pol} $, for SPC/E water  with the simplified approximations presented above, respectively, $\Delta G'_0 $ and $\Delta G_{{\rm pol}}^{\rm B} $. In order to shed further light on  the interplay between the solvent molecular dipole and quadrupole moments (and thus water
specificity), a series of molecular models having the same permanent dipole
moment as SPC/E, but different quadrupolar ones, were first generated by reducing
the H-O-H angle $\gamma $, while keeping fixed both the original partial
charges on each site and the distance between the oxygen and the midpoint
between hydrogen atoms.
{  The choice of including the variable quadrupole moment models was dictated by the need to find a smooth link via MD simulations between  a realistic water model (SPC/E) and the simplified purely dipolar models often studied (due to the inherent difficulty of  the problem) using approximate theoretical methods~\cite{28a,28}.  Due to symmetry the interfacial contributions under scrutiny here for liquids possessing molecular dipole and quadrupole moments must vanish for symmetric purely dipolar models.
We also would like to test the approximate
formulae for the quadrupolar contribution in a more general setting
(from SPC/E to a purely dipolar model) and to shed light on the coupling
between the dipolar and quadrupolar contributions.}
For all but one molecular model the SPC/E parameters were maintained for the Lennard-Jones (LJ) interaction centered on the oxygen atom. For the $n^{\rm th}$ molecule of each liquid model we define the molecular dipole moment $\mathbf p_n =\sum\nolimits_j q_j \mathbf{r}_{jn}$ and
quadrupole moment tensor
$(Q_n)_{\alpha \beta} = 3\sum\nolimits_j q_j x_{jn,\alpha} x_{jn,\beta}$,
with $x_{jn,\alpha}$ the $\alpha^{\rm th}$ Cartesian component of the position vector
$\mathbf r_{jn}$ of partial charge $q_j\; (j=1,2,3)$ with respect to the
center of charges within molecule $n$. We can then compute the macroscopic
polarization
    \begin{equation}
\mathbf{P}(\mathbf{r})= \left \langle \sum\nolimits_n \mathbf{p}_n \delta(\mathbf{r}-\mathbf{r}_n) \right \rangle
    \end{equation}
and the macroscopic quadrupole moment density,
    \begin{equation}
Q_{\alpha \beta }(\mathbf{r})=\frac1{6} \left \langle  \sum\nolimits_n (Q_n)_{\alpha \beta }  \delta(\mathbf{r}-\mathbf{r}_n) \right \rangle
    \end{equation}
 directly from the simulations as ensemble averages~\cite{2}.

The local electric charge density, $\rho(\mathbf{r})$, can be evaluated directly by
extracting the partial charge density associated with the particular molecular model;
and the associated electric field ${\bf{E}}$ and potential $\phi$ can then be obtained by
integration of the Poisson equation in appropriate coordinates:
    \begin{equation}
    \label{pe}
 \nabla ^2 \phi  = -\nabla  \cdot {\bf{E}} = -\frac{\rho }{{\varepsilon _0 }}.
    \end{equation}
The mean electric field along the direction normal to the planar vapor-liquid interface, at distance $z$
is written in Cartesian coordinates as:
    \begin{equation}
    \label{efz}
E_z \left( z \right) = \int\limits_{z_v }^z {\frac{{\rho \left( {z'} \right)}}{{\varepsilon _0 }}dz'},
    \end{equation}
where $\rho \left( {z'} \right)$
represents the average electric charge density (evaluated within the scope of MD simulations as the volume density of the sum of partial charges associated with a particular molecular model found in the bin corresponding to the position $z'$). The coordinate $z_v $ is  the origin of integration in the vapor phase (far from the interface).
Similarly, in spherical coordinates (appropriate for the curved solute-liquid interface), the mean electric field is given by
    \begin{equation}
    \label{efr}
E_r \left( r \right) = \frac{1}{r^2}\int\limits_{0 }^r {\frac{{\rho \left( {r'} \right) r'^2}}{{\varepsilon _0 }} dr'}.
    \end{equation}

The total charge density can also be expressed  as a multipole expansion~\cite{jack},
   \begin{equation}
\rho  =  - \nabla  \cdot {\bf{P}} + \sum_{\alpha, \beta} \nabla _\alpha  \nabla _\beta  Q_{\alpha \beta }  +  \ldots,
    \end{equation}
 here truncated after the  quadrupolar term. The dipolar and quadrupolar electric fields and potentials  can then be obtained from  the dipole moment density ${\bf{P}}$ and quadrupole moment density $Q$, respectively:
    \begin{equation}
    \label{ped}
  \nabla^2   \phi^D = -\nabla  \cdot {\bf{E}}^D   = \frac{1}{\varepsilon _0} \nabla  \cdot {\bf{P}} = -\frac{1}{\varepsilon _0} \rho^D
    \end{equation}
    \begin{equation}
     \label{peq}
  \nabla^2   \phi^Q = -  \nabla  \cdot {\bf{E}}^Q     = -\frac{1}{\varepsilon _0} \sum_{\alpha, \beta} \nabla _\alpha  \nabla _\beta  Q_{\alpha \beta } = -\frac{1}{\varepsilon _0} \rho^Q
    \end{equation}
After computing the full interface potential from eq.~\ref{pe}, we compare it with the sum of the dipolar and quadrupolar contributions computed from eqs.~\ref{ped} and \ref{peq} to assess the accuracy of the truncated multipole expansion.

The dipolar contribution to the total electric field can be obtained from $\rho ^D  =  - \nabla  \cdot {\bf{P}}$, the mean dipolar charge density created by the distribution of the macroscopic dipole moment density ${\bf{P}}$,  yielding for the $z$-component:
    \begin{equation}
     \label{efpz}
E_z^D \left( z \right) =  - \frac{{P_z \left( z \right)}}{{\varepsilon _0 }},
    \end{equation}
since $P_z$ vanishes at $z_v $ in the vapor phase, far from the interface region. Similarly, the dipolar component of the radial field in spherical coordinates, appropriate for the solute-liquid interface, reads:
    \begin{equation}
      \label{efpr}
E_r^D \left( r \right) =  - \frac{{P_r \left( r \right)}}{{\varepsilon _0 }},
    \end{equation}
where $P_r \left( r \right)$ represents the radial distribution of the density of the dipole moment. The dipole moment density ${\bf{P}}$ obtained from the MD simulations as ensemble averages of molecular dipole moments is presented in detail  below for the planar l-v interface (Section~\ref{dpo}).

The determination  of the mean electric field (or charge density) permits the  calculation of  $\delta \phi \left( z \right) = \phi \left( z \right) - \phi \left( {z_v } \right)$, the local electric potential difference evaluated at position $z$ in the vicinity of the interface:
    \begin{equation}
\delta \phi \left( z \right) = -\int\limits_{z_v }^z {E_z \left( {z'} \right)dz'} = {\frac{1}{{\varepsilon _0 }} \int\limits_{z_v }^z  {\rho \left( {z'} \right)\left( {z' - z} \right)} \, dz'}
    \end{equation}
Similarly, the dipolar local potential profiles, $\delta \phi ^D \left( z \right)$,  can be  obtained from the corresponding electric field, $E_z^D \left( z \right)$ (or charge density, $\rho^D$).

The quadrupole moments of the models SPC, SP9-SP5, and S2N range from the SPC/E
value down to zero (table~\ref{tab1}). Because both dipolar S2N and S2L models are
asymmetric, a symmetric dumbbell-like model (S2D) was also investigated
(with two LJ spheres on both ends of the dipole). Ions are modeled as simple
point charges carrying an LJ sphere.
\begin{table*}[t]
%% table 1
\caption{The system parameters [$\gamma$ is the H-O-H angle; $\mu^0$ and $Q^0_{xx}$ are the permanent molecular dipole and quadrupole moments; $\rho_l $ is the liquid density at the center
of the slab (estimated error of $\pm 0.005$~g/cm$^3$)], total interfacial potential $\phi_{lv}$, the corresponding quadru-  ($\phi _{lv}^Q $) and dipolar ($\phi _{lv}^D $) contributions, and the ``isotropic" quadrupolar approximation ($\phi _{lv,{\rm est}}^Q$, eq.~\ref{eq6}); Estimated standard error of $\pm 0.5$~mV for the vapor-liquid interface potentials of the studied liquid models (obtained using the block averaging method).}
\begin{center}
\begin{tabular}{|c|c|c|c|c|c|c|c|c|c|}

\hline

Model & $\gamma (^{\circ})$ &  $\mu^0$(D) & $Q_{xx}^0$(D{\AA}) & $\rho_l$(g/cm$^3$) & $\phi_{lv}$(mV) & $\phi_{lv} -\phi_{lv}^D$(mV) & $\phi_{lv}^Q$(mV) & $\phi _{lv,{\rm est}}^Q$(mV) & $\phi _{lv}^Q /\phi _{lv}$\\

\hline

SPC & 109.5 & 2.347& 8.131& 0.981& -600.3& -558.8& -559.2& -558.6& 0.932 \\

\hline

SP9& 100.0& 2.347& 5.775& 0.892& -445.8& -361.4& -360.1& -360.2& 0.808 \\

\hline

SP8& 87.6& 2.347& 3.736& 0.787& -254.8& -206.5& -206.8& -205.9& 0.811 \\

\hline

SP7& 75.0& 2.347& 2.394& 0.698& -123.6& -116.1& -116.9& -117.1& 0.946 \\

\hline

SP5& 54.7& 2.347& 1.089& 0.611& -21.4& -46.3& -46.2& -46.6& -- \\

\hline

S2N& 0& 2.347& 0& 0.556& 56.8& 0.2& 0& 0& 0 \\

\hline

S2L& 0& 4.065& 0& 0.696& 284.6& -0.4& 0& 0& 0 \\

\hline

S2D&0& 2.347& 0& 0.658& -0.8& -0.4& 0& 0& 0 \\

\hline
\end{tabular}
\label{tab1}
\end{center}
\end{table*}
Simulations of the vapor-liquid interface were carried out using a
modified parallel version of the molecular dynamics package Amber 9~\cite{amber} and a
slab geometry methodology similar to the one often used in the literature:
1000 liquid molecules placed in a rectangular unit cell of dimensions 31.04
$\times $ 31.04 $\times $ 91.04 {\AA}$^{3}$ occupying roughly the middle one-third of the available space and generating two vapor-liquid interfaces~\cite{24,24a}.
A Lennard-Jones interaction potential is centered on the solvant oxygen atom, characterized by the SPC/E parameters   $\sigma = 3.1657$~\AA\  and  $\varepsilon = 0.1553$~kcal/mol.
In the ``bulk" ion solvation simulations, the system comprised a cubic cell of 1000 water molecules, with one solute immersed at its center.

The ion properties  are summarized in table~\ref{tab2}. Each ion is modeled by a simple point charge, a Lennard-Jones potential defined by  $\sigma$  and   $\varepsilon$ and, when polarizable models are analyzed, a polarizability. The cross parameters for the ion-water Lennard-Jones interaction are determined via Lorentz-Berthelot mixing rules.
\begin{table*}[t]
%% table 2
\caption{Lennard-Jones parameters   $\sigma$  and   $\varepsilon$, and polarizability $\alpha$   for  ions}
\begin{center}
\begin{tabular}{|c|c|c|c|c|c|}

\hline

Ion	&  $q$ ($e$)  & $\sigma$ (\AA)  & $\varepsilon$  (kcal/mol)  & $\alpha$  & ref.   \\

\hline
Na${}^+$ 	& 1	& 2.350	& 0.13	& 0.24   & \cite{24a1}  \\

\hline

F${}^-$ &-1	&3.168	&0.2	&0.974	& \cite{24a2}   \\

\hline

Cl${}^-$ &	-1	&4.339	&0.1	&3.25&  	\cite{24a3}   \\

\hline

Br${}^-$	&-1	&4.700	&0.1	&4.53	&  \cite{24a4}   \\

\hline

I${}^-$ & -1	& 5.150	& 0.1	& 6.9	& \cite{24a4}   \\

\hline

\end{tabular}
\label{tab2}
\end{center}
\end{table*}

Periodic boundary conditions were applied in all three directions. The long-range charge-charge, charge-dipole and dipole-dipole interactions were treated by the particle-mesh Ewald summation method for both the charge and dipole moments~\cite{24b}.  For computational efficiency, in the polarizable simulations, an extended Lagrangian method was utilized to compute the induced dipole moments, regarded as additional dynamic variables~\cite{24c}.

A cutoff radius of 10~\AA\ was used for the short-ranged non-bonded LJ interactions and for the real space component of the Ewald summation. The geometries of the liquid molecules were constrained by applying the SHAKE algorithm  with a relative geometric tolerance of $10^{-4}$. The equations of motion were integrated using the velocity Verlet algorithm with a default time step of 1 fs~\cite{24d}. In order to avoid occasional drifts of the slab along the  \emph{z}-axis normal to the interface, the center-of-mass (COM) velocity was removed every 1000 steps. Configurations were saved every 100 fs in the output trajectories and each such frame was readjusted with respect to the  \emph{z}-axis, to keep the COM of the electrolyte fixed relative to the simulation cell.

Starting from the initial configuration of each simulated system, an energy minimization was performed, followed by a 1 ns NVT equilibration at 300 K for the slab systems. For simulations of bulk water ion solvation the equilibration process was performed in the NPT ensemble, using a weak-coupling pressure regulation with a target pressure of 1 bar. Subsequently, in both cases, at least 5 ns of measurements in the NVT ensemble were carried out, using the Berendsen thermostat with the configurational degrees of freedom coupled to a heat bath with coupling constant $\tau = 1$~ps~\cite{24e}.  In the special case of polarizability-enabled simulations, the degrees of freedom related to the induced dipole moment of the ion were independently coupled to a 1 K heat bath (relaxation time  $\tau_{\rm dip} = 10$~ps), ensuring a proper handling of the electronic degrees of freedom~\cite{24f}.
All computed profiles spanning the vapor-liquid interface were obtained as ensemble averages of the instantaneous profiles evaluated in thin slabs (bins) of thickness 0.2~\AA\
parallel to the interface. For the radial quantities measured in the bulk simulations of ion solvation, equally distanced, 0.1~\AA\ thick, spherical shell bins have been employed. Due to the cubic dimensions of the simulation box the radial profiles are relevant up to approximately 15~\AA.

\section{MD simulation results}
The MD simulation results show that the bulk region density
decreases with decreasing molecular quadrupole moment (at constant dipole
moment), varying by nearly a factor of two in going from SPC/E to the lowest density model, S2N (table~\ref{tab1}).
{   For the models possessing quadrupole moments,  SPC/E -- SP5, the density decreases by less than 40\%, despite a decrease by a factor of 8 in the molecular quadrupole moment. This density variation is an expected physical consequence of the reduction in water coordination  as the molecular quadrupole moment decreases.}
The S2N and S2L models form purely dipolar liquids with a characteristic
chain-like structure arising from the head-to-tail alignment of the dipole
moments and a lower bulk density due to the decrease in hydrogen bond
coordination from four to two. The vapor-water interfacial thickness is
found to be approximately 3.8~{\AA} at 300~K for SPC/E and increases along
with the slab thickness as the molecular quadrupole moment decreases.
    \begin{figure}
\resizebox{\columnwidth}{!}{\includegraphics{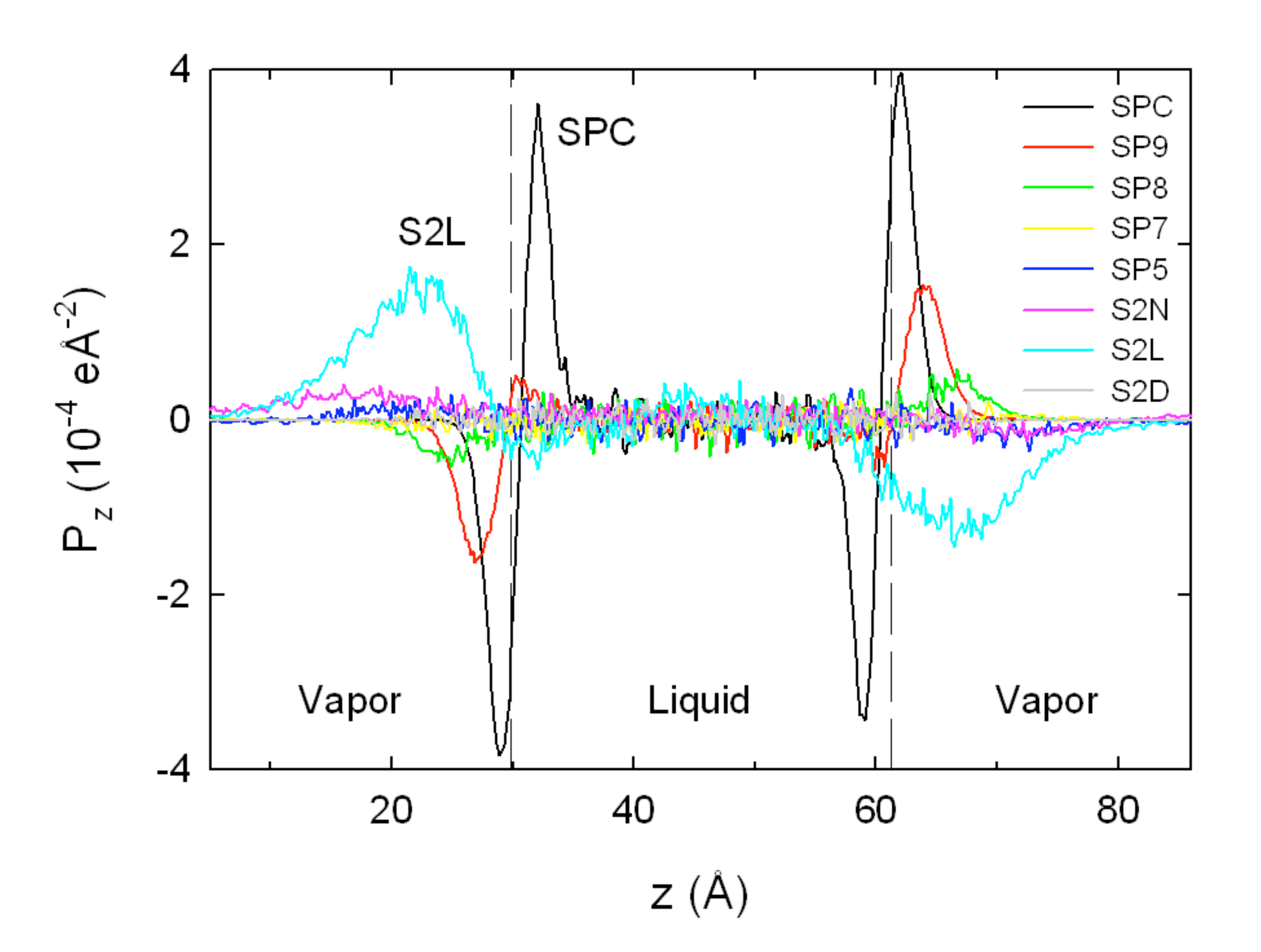}}
\caption{(Color on-line) The dipole moment density, $P_z
(z)$, for all studied liquids at the vapor-liquid interface. Dashed vertical lines represent the
GDS of both interfaces for SPC/E. }
\label{fig2}
    \end{figure}

\subsection{Dipolar ordering}
\label{dpo}
Orientational (dipolar) ordering of water takes place near the l-v interface,
which can be seen by plotting $P_z (z)$ for the series of model liquids
(fig.~\ref{fig2}). Orientational double layers were found only for SPC/E and SP9 with
the outer layer dipoles pointing preferentially towards the vapor phase and
in the opposite direction in the inner layers (closest to the slab center).
For models with lower molecular quadrupole moments the molecular dipoles
point towards the liquid phase. Because of the asymmetry created by the
oxygen LJ sphere, asymmetric purely dipolar liquids (S2N and S2L) still
possess orientational ordering in the interface region due to the
hydrophobic forces tending to exclude the oxygen LJ sphere from the liquid
slab.

\subsection{Vapor-liquid interface potential: multipole contributions} In order to illustrate how various multipole moments contribute to the
vapor-liquid interface potential, we obtained the electric potential
difference by integration of the Poisson equation from the charge density obtained from the first two terms of the multipole
expansion~\cite{2,10},
    \begin{equation}
\label{eq4}
\rho (z) \approx \rho^D(z)+\rho^Q(z) =  - \frac{d}{{dz}}\left[ {P_z \left( z \right) - \frac{d}{{dz}}Q_{zz} \left( z \right)} \right]
    \end{equation}
leading to the first two (dipolar and quadrupolar) contributions to the
interface potential:
    \begin{equation}
\label{eq5}
\delta \phi^{DQ}(z)\equiv \delta \phi
^D(z)+\delta \phi ^Q(z)=\int_{z_v }^z \frac{P_z
(z)}{\varepsilon_0} dz -\frac{Q_{zz} (z)}{\varepsilon _0}
    \end{equation}
since $Q_{zz} $ is taken to vanish in the vapor phase and $P_z $ vanishes in
both bulk (vapor and liquid) phases. For the planar interface higher order moments do not contribute.

\begin{figure}
\resizebox{\columnwidth}{!}{\includegraphics{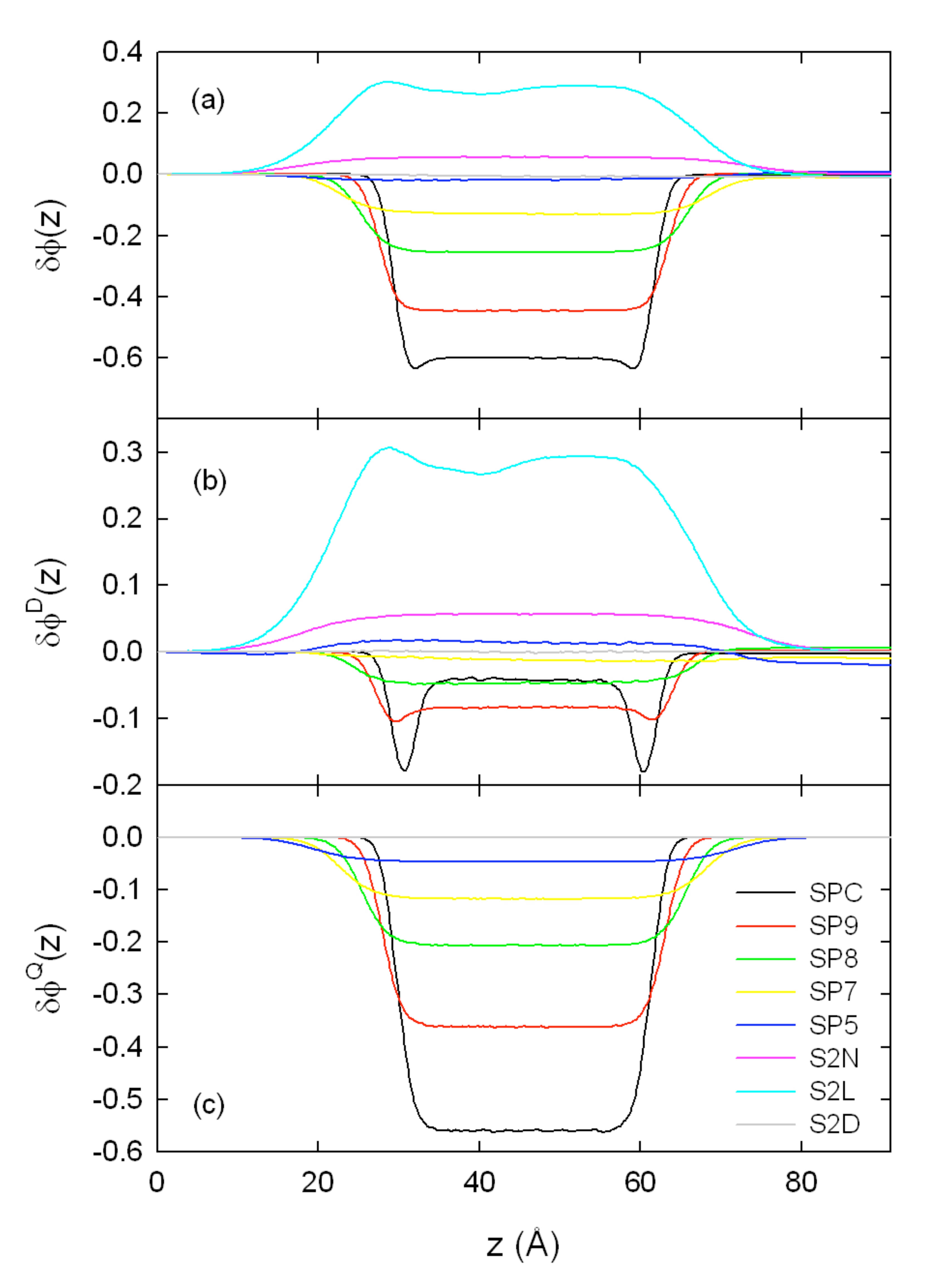}}
\caption{(Color on-line) (a) Total, (b) dipolar and (c) quadrupolar
potential profiles, $\delta \phi (z)$ (in volts) for the studied liquids characterized by positive molecular quadrupolar moments.}
\label{fig3}
    \end{figure}

\begin{figure}
\resizebox{\columnwidth}{!}{\includegraphics{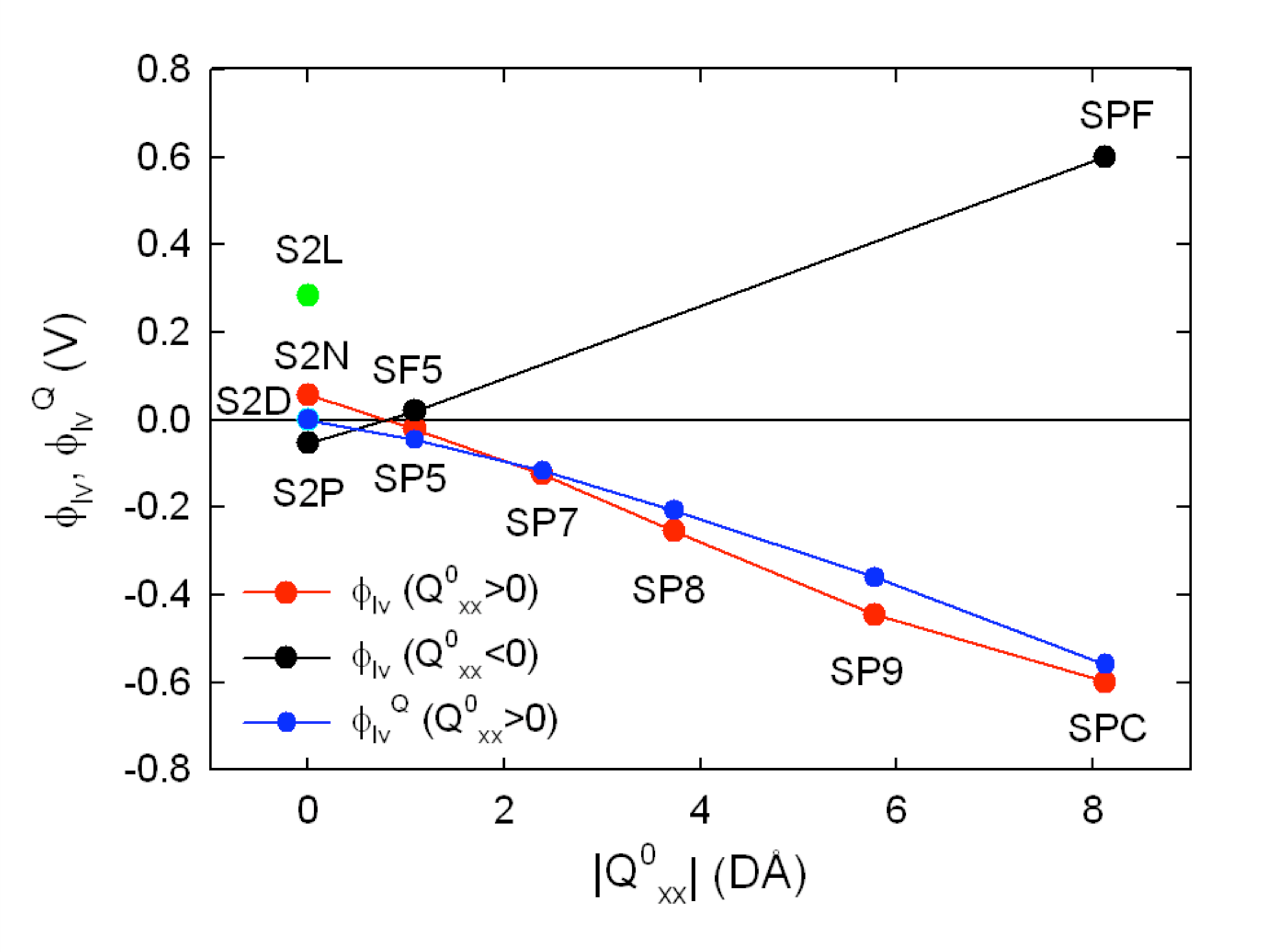}}
\caption{(Color on-line) Vapor-liquid interface potential drops for all
studied models: standard (blue) and flipped (F)-charge (black) with respect
to $|Q_{xx}^0|$ (molecular quadrupole moment).}
\label{fig4}
    \end{figure}

The ``exact'' model interface potential profile and the corresponding
dipolar and quadrupolar contributions (eq.~\ref{eq5}) obtained directly
from the simulation data (using, respectively, the ``exact'' partial charge
density, $\rho$, and the multipole contributions of eq.~\ref{eq4})  are
illustrated in fig.~\ref{fig3}. The total vapor-liquid potential reaches $-600$mV
for SPC/E, in agreement with previous values~\cite{5}, but decreases in amplitude
with decreasing molecular quadrupole moment (SP9 to SP5). The asymmetric
purely dipolar liquids, on the other hand, have positive interface
potentials, whereas, as expected by symmetry considerations, the fully symmetric model S2D gives a null result  (table~\ref{tab1}). The two components of the interface potential reveal
very different types of profiles with the quadrupolar potential being
negative, as expected for models with positive molecular quadrupole moments.
For the SPC/E model the quadrupolar contribution represents more than 90\%
of the total. This contribution decreases rapidly with
decreasing molecular quadrupolar moment and becomes comparable in absolute
value with the (positive) dipolar contribution for SP5, the near
cancellation in this case leading to a very low (negative) total value. For
SPC/E, SP9-SP7 the quadrupole contribution provides the major contribution
to the interface potential (figs.~\ref{fig3} and~\ref{fig4}, table~\ref{tab1}) and thus to the large
interfacial electric fields ($\sim1$~V/nm for SPC/E) directed towards the
liquid phase over a substantial part of the interfacial region~\cite{23}. For
SPC/E and SP9 the innermost dipoles tend to follow this field, creating in
turn an opposing dipolar field that acts to align the outermost molecular
dipoles in the opposite direction. Although the quadrupolar potential
profile is always monotonic, the dipolar one shows a minimum close to the
Gibbs dividing surface (GDS) for both SPC/E and SP9 models due to the
dipolar orientational bilayers. These results reveal a subtle interplay between the dominant quadrupolar contribution and the dipolar response.

Even if the system is not isotropic in the interfacial region, it is
possible to generalize the approach presented in~\cite{2,10,13} to construct a
simple but extremely accurate ``isotropic'' approximation for the
quadrupolar contribution to the local vapor-liquid interface potential using
only the water density profile and the molecular quadrupole
moment $Q^0$ evaluated in a local reference frame with the $y$-axis along the
dipole vector and the $z$-axis out of the molecular plane:
    \begin{equation}
\label{eq6}
\delta \phi_{\rm est}^Q (z)=-\frac{c(z)}{6\varepsilon _0}\frac{{\rm Tr}\;Q^0}3,
    \end{equation}
where $c(z)$ is the local liquid number density taken from the simulations.
 In this reference frame the only non-zero component is $Q^0_{xx}$ and therefore in this case ${\rm Tr}\;\textsf{Q}^0=Q_{xx}^0 $. The estimate $\phi _{lv,{\rm est}}^Q =\delta \phi _{\rm est}^Q (z_l)$
for the quadrupolar contribution is in excellent agreement with direct
determinations (table~\ref{tab1}).
{
We see from eq.~\ref{eq6} that the variations in vapor-liquid interface potentials for the models with non-zero quadrupole moments are mainly determined by the quadrupolar contribution and therefore dominated by the variations in the molecular quadrupole moments (with the physically relevant density variations playing only  a secondary role).  For this reason and  because the dipolar contribution is not strictly proportional to the liquid density, we do not attempt to normalize the vapor-liquid interface potentials in table~\ref{tab1} to correct for the variations in liquid density.}
We have also checked that this simple quadrupolar estimation \ref{eq6} provides a very good approximation to the full oscillatory membrane-water surface potential \cite{17}, confirming that the air-water interfacial and membrane-water surface potentials are mainly determined by the local water density and  molecular quadrupole moment.
Furthermore, we propose that this method can be used to estimate the quadrupolar potential contribution of any liquid,
irrespective of its bulk molecular quadrupole moment, even those obtained from ab initio quantum
mechanical calculations of liquid water. As an illustration, we have checked that when ab initio values for ${\rm Tr}\;\textsf{Q}^0$~\cite{29a,24g}, are injected into the simple approximation Eq.~\ref{eq6}, we find quadrupolar potentials in  reasonable agreement with the (quadrupole dominated) total interface potentials (+3~eV) extracted directly from the quantum mechanical calculations~\cite{12,13}.

\subsection{Solute-Liquid interface potential}
In the presence of a solute the planar vapor-liquid interface potential has
as counterpart the microscopic potential between the solute center and the
surrounding liquid (of which the first two multipole terms can be obtained
from the microscopic analog of eq.~\ref{eq5}~\cite{23}).

The local radial solute-liquid potential profile, defined as
    \begin{equation}
\delta \phi _r \left( r \right) = \phi _r \left( r \right) - \phi _r \left( {r_s } \right)
    \end{equation}
[with $\phi _r \left( {r_s } \right) = 0$
at the center of the solute $r_s = 0$], is obtained from the radial charge density
$\rho _r \left( r \right)$ by integrating the Poisson equation (\ref{pe}) in spherical coordinates:
    \begin{equation}
    \label{potr}
\delta \phi _r \left( r \right) =  - \int\limits_0^r  E_r (r')  dr' = - \int\limits_0^r {\frac{{\rho _r \left( {r'} \right)\left( {r'} \right)^2 }}{{\varepsilon _0 }}\left( {\frac{1}{{r'}} - \frac{1}{r}} \right) dr'}.
    \end{equation}
The dipolar component, $\delta \phi _r^D \left( r \right)$, is determined from the dipole moment density $P_r \left( r \right)$  as:
    \begin{equation}
      \label{potrd}
\delta \phi _r^D \left( r \right) = \int\limits_0^r {\frac{{P_r \left( {r'} \right)}}{{\varepsilon _0 }}dr'}
    \end{equation}
The corresponding radial electric fields, $E_r \left( r \right)$ and $E_r^D \left( r \right)$, are determined from equations (\ref{efr}) and (\ref{efpr}).

The quadrupolar contribution to the total radial interface potential, $\delta \phi _r^Q \left( r \right)$ is accessible from the simulation data via the radial dependence of the quadrupole moment density written in spherical coordinates, $\textsf{Q}' (r )$. We begin by writing the quadrupolar Poisson equation (\ref{peq}) in Cartesian coordinates (centered at the solute position $r_s  = 0$), with the tensor elements of the Cartesian quadrupole moment density, $Q_{\alpha \beta } $:
    \begin{equation}
\varepsilon _0 \nabla  \cdot {\bf{E}}^Q  = \sum\limits_{\alpha ,\beta } {\nabla _\alpha  \nabla _\beta  Q_{\alpha \beta } }
    \end{equation}
Using the divergence theorem, we find:
    \begin{equation}
\varepsilon _0 \mathop{{\int\!\!\!\!\!\int}\mkern-21mu \bigcirc}\limits_S
 {\bf{E}^Q  \cdot d{\bf{S}}}  = \mathop{{\int\!\!\!\!\!\int}\mkern-21mu \bigcirc}\limits_S
 {\sum\limits_{\alpha ,\beta } {\nabla _\alpha    Q_{\alpha \beta } }  \cdot n_\beta  dS},
    \end{equation}
{   where $\hat{\bf{n}}$ is the normal to the interface.}
Letting $F_\beta   = \sum\limits_\alpha  {\nabla _\alpha  Q_{\alpha \beta } } $ and $S = 4\pi r^2 $, we obtain the radial dependence of the quadrupolar electric field $E_r^Q \left( r \right)$, by integrating over the angular degrees of freedom:
    \begin{equation}
4\pi \varepsilon _0 E_r^Q \left( r \right) = \mathop{{\int\!\!\!\!\!\int}\mkern-21mu \bigcirc}\limits_S
 {F_r \sin \theta d\theta d\phi },
    \end{equation}
with $F_r  = \bf{e}_r  \cdot \bf{F}$
{  (where $\bf{e}_r$ is the unit radial vector).}
After performing the angular integrals, we obtain the radial dependence of the quadrupolar electric field:
    \begin{equation}
\varepsilon _0 E_r^Q \left( r \right) = \frac{{\partial Q'_{rr} }}{{\partial r}} + \frac{1}{r}\left( {3Q'_{rr}  - TrQ} \right).
    \end{equation}
Since $E_r^Q \left( r \right) =  - \frac{{\partial \phi_r^Q \left( r \right)}}{{\partial r}}$ and $\phi_r^Q \left( 0 \right) = 0$, we arrive at  the  final solution for the local quadrupolar potential at position $r$, with respect to the center of the solute, $\delta \phi _r^Q  = \phi _r^Q \left( r \right) - \phi _r^Q \left( {r_s } \right)$,  in terms of two components:
    \begin{equation}
    \label{potrq}
\delta \phi _r^Q  (r) = \delta \phi _r^{Q,1} (r) + \delta \phi _r^{Q,2} (r)
    \end{equation}
with
    \begin{eqnarray}
  \delta \phi _r^{Q,1} \left( r \right) &=& - \frac1{\varepsilon _0 } Q'_{rr} \left( r \right) \label{potrq1} \\
  \delta \phi _r^{Q,2} \left( r \right)&=& \frac1{\varepsilon _0 } \int\limits_0^r \frac{dr' }{{r'}} \left[ {\rm Tr} \textsf{Q}({r'}) - 3Q'_{rr} ({r'}) \right].
  \label{potrq2}
    \end{eqnarray}

The second contribution  is generated by the symmetry breaking of the diagonal components of $\textsf{Q}'$ in the solute-liquid interfacial region.
The detailed calculations leading from $\textsf{Q}\left( {\bf{r}} \right)$ to
$\delta \phi _r^Q \left( r \right)$ will be presented elsewhere~\cite{23}.  Far from the solute center the radial solute-liquid interface potential and the  various multipole components tend to their respective asymptotic values: $\phi _{ls} = \delta \phi _r (\infty)$,  $\phi _{ls}^D = \delta \phi _r^D  (\infty)$, and $\phi _{ls}^Q = \delta \phi _r^Q  (\infty)$.

Because the solute-liquid
interface is curved, the associated potential depends on the size of the
cavity and is therefore ion specific. Thus, there is a first potential drop
when going from the vapor into the liquid phase, followed by an overall
increase near the solute, yielding a smaller, overall negative,
vapor-liquid-solute potential drop, $\phi _{sv} $ (fig.~\ref{fig1}). To obtain a
fuller picture, we have extracted the solute-liquid interfacial potentials (and the dipolar/quadrupolar components)
from MD simulations of neutral ion-like solutes, fixed at the center of a
cubic box of bulk SPC/E water (using eqs.~\ref{potr}, \ref{potrd},and \ref{potrq}).
 The solute-liquid potential $\phi _{ls}  $
and the corresponding dipolar contribution $\phi _{ls}^D $ are obtained,
respectively, from the radial charge density and the dipole moment density
distributions.
The higher order multipole contributions, $\phi _{ls} -\phi _{ls}^D $, are dominant and, due to \textit{ls} interface
curvature effects, not only is the amplitude of the \textit{ls} quadrupolar
contribution different from the \textit{lv} one, but multipole terms beyond the quadrupolar one play a non-negligible role (tables~\ref{tab3} and \ref{tab4})~\cite{23}.
For the halide-like neutral solutes $\phi _{ls}$ decreases in amplitude with increasing solute size and is smaller than for the neutral sodium like solute, Na${}^0$.  For I${}^0$,  $\phi _{ls}$ increases in amplitude  by about 6\% when the polarizability is turned on.
Although the $\phi _{ls}^{Q,1}$ contribution is dominant in $\phi _{ls}$,  it simply serves, as we shall see below,  to cancel the large vapor-liquid quadrupolar contribution to  the \textit{direct} contribution to the free energy of ion solvation, $\Delta G_0$.
 The dipole contribution $\phi _{ls}^D$ goes from negative values for small halide-like solutes (F${}^0$) to positive values for larger ones.

\subsection{Free energy of ion solvation}
The solute-liquid potential was then used to
evaluate the \textit{direct} contribution to the free energy of ion solvation,
$\Delta G_0 = q(\phi_{lv} -\phi _{ls})$ (eq.~\ref{eq1a}),  dominated by the difference between the quadrupolar \textit{lv} and \textit{ls} contributions, which do not cancel due to strong curvature effects for the small ions under study.  Because the $\phi _{ls}^{Q,1}$ component depends only on the bulk solvent properties, it  is in principle equal to $\phi _{lv}^{Q}$ and therefore the two terms should cancel in $\Delta G_0$, leaving the dominant  ion specific  quadrupolar contribution, $\phi _{ls}^{Q,2}$ (tables~\ref{tab3} and \ref{tab4}).
The dipolar contribution and  multipolar contributions higher than quadrupolar  play non-negligible, but secondary roles, in $\Delta G_0$.  Our results for polarizable $I^{-}$ also reveal that ionic polarizability plays a minor but non-negligible role in determining $\Delta G_0$ (tables~\ref{tab3} and \ref{tab4}).
%%%%%%%
\begin{table*}[ht!]
%% table 3
\caption{Solute-liquid interface potentials with multipolar contributions for neutral solutes solvated in SPC/E water (for which the vapor-liquid interface potential is  $\phi_{lv} = -600.3$~mV with  $\phi_{lv}^Q = -559.2$~mV and   $\phi _{lv}^D = -41.5$~mV).  In principle  $\phi_{ls}^{Q,1} = \phi_{lv}^Q$. The less than 2\% difference between the  values obtained from the water slab and bulk simulations can be attributed to the differences in densities generated by the finite slab thickness  and the barostat used in the NPT bulk simulations (cf. eq~\ref{eq6}).  To directly compare solute-liquid and vapor-liquid interface potentials, we  correct for this small systematic liquid density difference by defining  a  solute-dependent  rescaled  vapor-liquid interface potential, $\phi_{lv}^* = C \phi_{lv}$, where  $C \equiv \phi_{ls}^{Q,1}/\phi_{lv}^Q$  (rescaled  multipole vapor-liquid interface components are defined similarly). The estimated standard error for the interface potentials is ±0.15~mV and \textit{pol} denotes polarizable. }

\begin{center}
\begin{tabular}{|c|c|c|c|c|c|c|c|c|}

\hline

Solute& $\phi_{ls}/\phi_{lv}^*$ (\%)& $\phi_{ls}$(mV) & $\phi_{ls}^D$ (mV)& $\phi_{ls}^Q$ (mV)& $\phi_{ls}^{Q,1}$ (mV) & $\phi_{ls}^{Q,2}$ (mV) ) & $\phi_{ls}-\phi_{ls}^D-\phi_{ls}^Q$(mV)  \\

\hline

Na$^{0}$ & 66.7  & 	-405.7	& -78.8	& -423.6	& -566.7	& 143.1	& 96.7 \\

\hline

F$^{0}$& 62.4&	-380.0&	-41.6&	-415.3&	-567.3&	152.0&	76.9 \\

\hline

Cl$^{0}$& 60.3& -369.0&	0.3&	-426.8&	-569.9&	143.1&	57.5 \\

\hline

Br$^{0}$& 58.8&	-359.8&	17.8&	-430.8&	-569.6&	138.8&	53.2 \\

\hline

I$^{0}$& 58.4&	-357.5&	 32.3&	-437.6&	-570.5&	132.9&	47.8\\

\hline

I$^{0}$ (pol) & 62.1&	-380.1&	14.4 &	-440.0& 	-570.5& 	130.5&	45.5
 \\

\hline

\end{tabular}
\label{tab3}
\end{center}
\end{table*}
\begin{table*}[ht!]
%%% Table 4
\caption{Comparison of the different multipole contributions to the \textit{direct} electrostatic solvation-free energies, $\Delta G_0^* = q(\phi_{lv}^*-\phi_{ls})$, obtained using the rescaled vapor-liquid interface potentials (*), of  various ions as obtained directly from the simulations (see table~\ref{tab2}) ($\Delta {G}'_0 \simeq \pm 0.6003$~eV);  estimated errors: 0.002~eV}.
\begin{center}
\begin{tabular}{|c|c|c|c|c|}

\hline

Solute &   $\Delta G_0^*$ (eV)& Dipole${}^*$   (\%)  & Quadrupole${}^*$ (\%) &  Higher order${}^*$ (\%)\\

\hline

Na$^{+}$ & -0.2030 & -18.1 & 70.5 & 47.6 \\

\hline

F$^{-}$& 0.2294 & 0.2 & 66.3 & 33.5 \\

\hline

Cl$^{-}$& 0.2432 & 17.5 & 58.8  & 23.6 \\

\hline

Br$^{-}$& 0.2521 & 23.8  & 55.1 & 21.1 \\

\hline

I$^{-}$& 0.2554 & 29.2  & 52.0 & 18.8 \\

\hline

I$^{-}$ (pol) & 0.2327& 24.4 & 56.0 & 19.6 \\

\hline

\end{tabular}
\label{tab4}
\end{center}
\end{table*}
\begin{table*}[ht!]
%%% Table 5
\caption{Comparison of electrostatic solvation-free energies for various
ions as obtained directly from the simulations, $\Delta G_{{\rm ES}}^{{\rm ion}} =\Delta G_0 +\Delta G_{{\rm pol}}$, with the
estimate $\Delta G_{{\rm ES,est}}^{{\rm ion}} =\Delta {G}'_0 +\Delta
G_{{\rm pol}}^{\rm B}$ obtained from the truncated direct contribution
[$\Delta {G}'_0 \simeq \pm 0.6$~eV for monovalent anions (resp. cations)] and the Born approximation, $\Delta
G_{\rm pol}^{\rm B}$ . $R_i$ is chosen as the distance from the ion center where the ion-water radial
distribution functions first reach 1~\cite{23}. To correct for the small systematic differences in  liquid density  between water slab and bulk simulations, the rescaled vapor-liquid interface potentials, $\phi_{lv}^*$, are used in the evaluation of the direct contribution: $\Delta G_0^* = q (\phi_{lv}^*-\phi_{ls})$ (see table~\ref{tab3}).
Estimated  error of  0.025{\AA} for $R_i $; Estimated  standard errors: 0.002~eV for $\Delta
G_0 $; 0.05~eV for the other solvation-free energies. (1~eV$=23.06$~kcal/mol$\simeq 40 k_{\rm B}T$)}
\begin{center}
\begin{tabular}{|c|c|c|c|c|c|c|c|}

\hline

Solute& $R_i$ (\AA) & $\Delta G_0^*$ (eV)& $\Delta G_{\rm pol}^{\rm B}$ (eV) & $\Delta G_{\rm pol}$ (eV) & $\Delta G_{\rm ES,est}^{\rm ion}$ (eV) & $\Delta G_{\rm ES}^{\rm ion}$ (eV) & $\Delta G_{\rm exp}^{\rm ion}$ (eV)~\cite{29a}\\

\hline

Na$^{-}$ & 1.05& 0.2030& -6.76& -6.88& -6.57& -6.68& -- \\

\hline

F$^{-}$& 1.55& 0.2294& -4.58& -4.46& -4.36& -4.23& -4.50 \\

\hline

Cl$^{-}$& 2.00& 0.2432& -3.55& -3.34& -3.32& -3.10& -3.64 \\

\hline

Br$^{-}$& 2.25& 0.2521& -3.15& -3.01& -2.91& -2.76& -3.30 \\

\hline

I$^{-}$& 2.45& 0.2554& -2.90& -2.63& -2.66& -2.38& -2.90 \\

\hline

Na$^{+}$& 2.20& -0.2030& -3.23& -3.88& -3.42& -4.08& -4.26 \\

\hline

I$^{+}$ & 3.55& -0.2554& -2.00& -1.79& -2.24& -2.04& -- \\

\hline

\end{tabular}
\label{tab5}
\end{center}
\end{table*}
The key point is that $\Delta G_0$  is much smaller in amplitude than the simple direct estimate ($|\Delta G'_0| = 0.600$~eV)   because  of  the strong partial  cancellation of the two interface potentials, $\phi_{lv}$ and $\phi _{ls}$.
Our results show that the direct contribution, $\Delta G_0 $ is reduced to
30-40\% of the simple direct estimate $\Delta G'_0 $ and depends on the sign of the ion charge in such a way as to favor the solvation of cations Na$^{+}$ with respect to the anions
(halogens).  For the halogens $\Delta G_0 $ is positive  and increases with increasing ion size, thus augmenting the interface propensity  of  large anions like I${}^-$.
Because the amplitude of $\Delta G_0 $ ($\sim\, 0.20 - 0.26$~eV) is still
between 5\% (for F${}^-$) and 10\% (for I${}^-$) of the dominant polarization contribution, $\Delta
G_{\rm pol}$ (see table~\ref{tab5}), it is on par with other important
contributions (such as the hydrophobic one~\cite{16,18,18a,19a,25}) and therefore must be
taken into account correctly when considering ion-specific effects.

Because of charge-dipole and charge-quadrupole coupling, when
the ion charge is ``turned on'' the solvent molecules  reorient
themselves in order to accommodate the solute, optimizing as much as
possible the orientation and number of hydrogen bonds: the ion polarizes the
medium around it and induces a radial potential difference, $\phi_{ls}^{\rm ion} $, dominated by the ``long range'' dipolar contribution.
The difference between $\phi _{ls}^{\rm ion} $ and $\phi _{ls} $
extracted from the simulations then determines $\Delta G_{\rm pol}$ (eq.~\ref{eq1b}). With our choice of effective ion radius, the simple mesoscopic Born approximation $\Delta G_{\rm pol}^{\rm B} $ is in
reasonably good agreement with the microscopic polarization contribution, $\Delta G_{\rm pol}$  (table~\ref{tab5}).  The  polarization contribution to the solvation free energy, which is the main effect  favoring high ion solvation, decreases in amplitude with increasing halogen size. We note also that the simulation results for the electrostatic contribution to the ionic free energy of solvation, $\Delta G_{\rm ES}^{\rm ion}$, follow  the experimental trend, $\Delta G_{\rm exp}^{\rm ion}$, reasonably well (table~\ref{tab5}), despite the neglect of certain entropic (hydrophobic) and enthalpic contributions (arising, for example,  from the short range repulsion and the long range van der Waals---dispersion---attraction).

\section{Conclusion}
We have studied a series of liquid models
that interpolate between SPC/E water and pure dipolar liquids and shown that
the quadrupolar component of the vapor-liquid interfacial potential typically dominates for the studied liquids
possessing a non-zero quadrupolar moment. In an effort to elucidate the
different ion-specific contributions to the free energy of solvation, we
have shed light on the key role played by the solute-liquid interface
potential and demonstrated that it leads to a strong reduction in the direct
electrostatic contribution with respect to previous estimates based solely on the
vapor-liquid potential.

We propose that the same mechanism would be at play
if the \textit{point} partial charge distribution of the solvent extracted from classical
MD simulations were replaced by the more realistic \textit{extended} charge distributions
found in \textit{ab initio} calculations.
Indeed, a coarse graining procedure for the electric
potential proposed recently~\cite{13}, which corrects for regions inaccessible to
ionic probes, shows that, encouragingly, both \textit{ab initio} and point charge
coarse grained potentials converge to values that are compatible with the
results for $\Delta G_0 =q\phi _{sv}$ presented in table~\ref{tab5}.
Finally the
dominant electrostatic polarization contribution to the free energy of
solvation was found to agree reasonably well with a Born-type approximation.
We conclude that the \textit{direct} interface potential contribution to the ionic free energy of solvation (or PMF) can neither be estimated using the point ion approximation (leading to a gross overestimate), nor be neglected entirely -- the two approximations commonly adopted in the current literature.
An  important corollary  that can be drawn from our study is that, in contradistinction to what is sometimes suggested~\cite{10}, even purely quadrupolar liquids should  give rise to an interface potential contribution to the ionic solvation free energy because of  the incomplete cancellation of the \textit{lv} and \textit{ls} components (due to solute curvature effects).
{   The mechanism investigated here leading to a strongly reduced vapor-liquid interfacial potential contribution to the electrostatic part of the ionic PMF is quite general and should be applicable not only to membrane-liquid surfaces, but also  other types of solvents and solutes. More complicated, possibly  non-spherically symmetric,  ions---like large organic ones---can be built for MD simulations from several charged LJ particles and therefore an ionic cavity devoid of solvent will form and give rise to a solute-liquid interface potential.}

{ It would also be interesting, albeit difficult, to generalize the approximate theoretical statistical mechanical approaches developed previously for
dipolar liquid models  near interfaces~\cite{28a,28} to quadrupolar liquids  in order to capture
the effects studied here via MD simulations.}
An important outcome of a reliable  mesoscopic theoretical approach to the problem investigated here would be a greatly enhanced comprehension of the underlying physics of ion
distributions in inhomogeneous dielectric settings with important applications in colloidal science,
nanotechnology (ion transport in artificial nano-pores, or nano-filtration), and biophysics (ion channels, biological membranes, DNA)~\cite{14a,14b,14c,14d}. We also expect that the results presented here transcend the particular chosen models and thus qualitatively illustrate important physicochemical mechanisms at play in ion partitioning within inhomogeneous dielectric media.

{   After this work was completed we became aware of other interesting very recent work covering similar topics and in which some of the same conclusions were reached \cite{29,30,31,32}. In \cite{29}  the same problem was studied and the same interface potential reduction mechanism proposed from a different perspective:  instead of  investigating the various multipole contributions, as we do, a novel  method of partitioning the ionic solvation free energy was used to extract from MD simulations of SPC/E water  different physically identifiable (cavity formation, attractive van der Waals, local  and far-field electrostatic) contributions (using recently optimized MD parameters~\cite{19a,30a}). Interestingly, for  I${}^-$,  the loss of the first water hydration layer (local electrostatic contribution), favoring ion solvation, is nearly counterbalanced by the hydrophobic  (cavity formation) contribution favoring desolvation. A remaining net interface potential contribution, favoring anion desolvation,  due to the competing vapor-liquid and solute-liquid interfaces, was obtained that is consistent with the results obtained here.  The authors of \cite{29} also proposed, as we did above, that this same interface potential reduction mechanism could be used to  resolve the apparent huge discrepancy between the classical and quantum predictions for the interface potential contribution to the ionic solvation free energy. This scenario was shown to be  viable  in \cite{32}, where both a detailed critical comparison with other quantum simulations and a  favorable experimental assessment were carried out.
The results obtained in \cite{32} are consistent with those of \cite{13} showing that a coarse graining procedure that effectively omits  certain regions of space in computing average interface potentials leads to a closer agreement between  the classical and quantum results. Despite these recent advances concerning the interface potential contribution to the ionic solvation free energy, a definite comparison with experiment is for the moment complicated by what appears to be some model and/or sampling dependency.
In \cite{30}  the free energy of a single ion close to hydrophobic
and hydrophilic surfaces was investigated using a novel  theoretical framework to
obtain the position dependent dielectric response  for interfacial water using molecular dynamics simulations. A multipole analysis for the planar surface was then carried out, underlining the importance of the quadrupolar contribution to this surface potential (consistent with the results presented here). The role of the solute-liquid interface contribution was not, however,  evoked in \cite{30}.  In \cite{31} the driving forces for anion adsorption to the water vapor-liquid  interface were studied  by comparing the results of MD simulations with those of a simplified mesoscopic theoretical  approach. Two different solvent models were used, SPC/E water and a symmetric purely dipolar liquid (the Stockmayer model, similar to our S2D model, presented above). On the one hand, an  extra electrochemical surface potential contribution was needed in the mesoscopic theoretical  approach  to explain the results of the  SPC/E simulations and the magnitude and sign of this contribution are consistent with the reduced  electrostatic interfacial one obtained here. On the other hand, no such extra electrochemical contribution was needed in the case of the Stockmayer model, which  is consistent with our result that there should be no interfacial electrostatic contribution for symmetric purely dipolar liquids.}

The simulation results presented here should be not only a useful building
block in the quest for constructing the physically relevant mesoscopic components of
the ion free energies of solvation (or PMF), but also provide a challenge for
statistical mechanical approaches used to analyze the subtle interplay
between short range steric, dipolar, and quadrupolar interactions in
determining the properties of polar fluids (in particular the dipole
distribution near interfaces) and their influence on ions.

We are currently extending the approach developed here for the ionic solvation
free-energy to the study of the local ionic PMF near aqueous interfaces and
surfaces with and without the potentially important effect of water and ion
polarizability~\cite{17,18,18a,24,25,26}.
{     We note that unlike certain non-polarizable ion-water models \cite{19a,30a},  polarizable ones have not yet been properly optimized and therefore the relative weight of polarizability in driving large anions to interfaces and surfaces (compared with the interfacial electrostatic contribution investigated here) remains to be determined.}

\begin{acknowledgments}
We acknowledge financial support from the French Agence universitaire de la francophonie (AUF) and ANR (grant ANR-07-NANO-055-01, SIMONANOMEM project) and the Romanian CNCSIS PN-II 502 and 506 grants.
This work is  partly based on an ANR SIMONANOMEM project report posted on the web in 2010
{\footnotesize \verb"http://www.lpt.ups-tlse.fr/spip.php?article718&lang=en"} and on the Ph.D. thesis of Lorand Horvath.  We would also  like  to  thank B.  Coasne for helpful discussions and T. Beck,   S. Kathmann,  K. Leung, and R. Netz for communicating their work to us. We are also grateful to  R. Netz for helpful comments on an earlier version of the  manuscript.

\end{acknowledgments}

%%%%%%%%% biblio %%%%%%%%%%%%


\begin{thebibliography}{0}
\bibitem{1} F. H. Stillinger and  A. Ben-Naim  {J. Chem. Phys.}\textbf{47}, 4431 (1967).
\bibitem{2} M. A. Wilson, A. Pohorille,  and L. R. Pratt, {J. Phys. Chem.} \textbf{91},  4873 (1987).
\bibitem{3} M. A. Wilson, A. Pohorille,  and L. R. Pratt, {J. Chem. Phys.} \textbf{88},  3281 (1988).
\bibitem{4} M. A. Wilson, A. Pohorille,  and L. R. Pratt, {J. Chem. Phys.} \textbf{90},  5211 (1989).
\bibitem{5} V. P. Sokhan and D. J. Tildesley, {Mol. Phys.} \textbf{92},  625 (1997).
\bibitem{6} L. X. Dang and T.-M. Chang, {J. Phys. Chem. B}\textbf{106},  235 (2002).
\bibitem{7} E. N.Brodskaya and V. V. Zakharov, {J. Chem. Phys.} \textbf{102},  4595 (1995).
\bibitem{8} M. Paluch,  {Adv. Colloid Interface Sci.} \textbf{84},  27 (2000).
\bibitem{9} M. Matsumoto and Y.  Kataoka , {J. Chem. Phys.} \textbf{88},  3233 (1988).
\bibitem{10} E. Harder and B. Roux, {J. Chem. Phys.} \textbf{129},  234706 (2008).
\bibitem{10a} {   I. Vorobyov and T.  W. Allen,  {J. Chem. Phys.} \textbf{132},  185101 (2010).}
\bibitem{11} S. M.Kathmann, I.-F. W. Kuo,  and C. J. Mundy,  {J. Am. Chem. Soc.} \textbf{131},  17522 (2008).
\bibitem{12} K. Leung,   {J. Phys. Chem. Lett.} \textbf{1},  496 (2010).
\bibitem{13} S. M. Kathmann . W. Kuo, C. J. Mundy, and G. K. Schenter, {J. Phys. Chem. B.} \textbf{115},  4369 (2011).
\bibitem{14} M. D. Baer and C.J. Mundy,  {J. Phys. Chem. Lett.} \textbf{2},  1088 (2011).
\bibitem{14a} W. Kunz, {Specific ion effects} (World Scientific Publishing, Singapore, 2010).
\bibitem{14b} P.  H\"{u}nenberger,  M.  Reif, {Single-Ion Solvation: Experimental and Theoretical
    Approaches to Elusive Thermodynamic Quantities} (Royal Society of Chemistry , Cambridge, UK, 2011).
\bibitem{14c} 	J. N. Israelachvili, {Intermolecular and Surface Forces}, 2nd edition, (Academic Press, London, 1992).
\bibitem{14d}  R. J. Hunter {Foundations of Colloid Science} (Oxford Universtiy Press, Oxford, 2002).
\bibitem{15} M. Bostr\"{o}m,  W. Kunz,  and B. W. Ninham, {Langmuir} \textbf{21}, 2619 (2005).
\bibitem{17} D. Horinek and R. R. Netz, {Phys. Rev. Lett.} \textbf{99},  226104 (2007).
\bibitem{16} D. M.Huang, C. Cottin-Bizonne , C. Ybert,  and L. Bocquet, {Langmuir} \textbf{24}, 1442 (2008).
\bibitem{18} Y. Levin, {Phys. Rev. Lett. }\textbf{102},  147803 (2009).
\bibitem{18a} Y. Levin, A.P. dos Santos,  and A. Diehl, {Phys. Rev. Lett.}\textbf{103},  257802 (2009).
\bibitem{19} M. Manciu and E. Ruckenstein,  {Langmuir }\textbf{21},  11312 (2005).
\bibitem{19a} D.  Horinek, A. Herz, L. Vrbka, F. Sedlmeier, S. I. Mamatkulov, and
R. R. Netz, {Chemical Physics Letters} \textbf{479}  173 (2009).
\bibitem{20} S. Buyukdagli,  M. Manghi,  and J. Palmeri, {Phys. Rev. Lett.} \textbf{105}, 158103 (2010).
\bibitem{spce}   H. J. C. Berendsen, J. R. Grigera,  and T. P. Straatsma, {J. Chem. Phys.} \textbf{91}, 6269 (1987).
\bibitem{22} D. A. McQuarrie, {Statistical Mechanics,} 2$^{nd}$ edition (University Science Books, Sausalito, California, 2000), Chap.15.
\bibitem{23} L. Horv\'{a}th, T. A. Beu,  M. Manghi, and J. Palmeri, {in preparation.}
\bibitem{28a}  Peter Frodl and  S. Dietrich, {Phys. Rev. E} \textbf{48},  3741 (1993).
\bibitem{28}  A. Abrashkin, D. Andelman,  and H. Orland, {Phys. Rev. Lett.} \textbf{99},  077801 (2007).
\bibitem{jack} J. D. Jackson, {Classical Electrodynamics}, 3$^{rd}$ edition (John Wiley \& Sons, New York, 1999).
\bibitem{amber} D. A. Case {et al.}, AMBER 9 (University of California, San Francisco, 2006).
\bibitem{24} P.  Jungwirth and  D. J. Tobias, {Chem. Rev.} \textbf{106},   1259 (2006).
\bibitem{24a} T.-M. Chang and  L. X. Dang, {Chem. Rev.} \textbf{106},   1305 (2006).
\bibitem{24a1}  D. E. Smith and L. X. Dang, {J. Chem. Phys.} \textbf{100}, 3757 (1994).
\bibitem{24a2}  L. Perera and M. L. Berkowitz, {J. Chem. Phys.} \textbf{100}, 3085 (1994).
\bibitem{24a3}  L. X. Dang, {J. Phys. Chem. B} \textbf{106}, 10388 (2002).
\bibitem{24a4} G. Markovich, L. Perera, M. L. Berkowitz,  and O. Cheshnovsky,{ J. Chem. Phys.} \textbf{105}, 2675 (1996).
\bibitem{24b} T. Darden, D. York,  and L. Pedersen, { J. Chem. Phys.}  \textbf{98},  10089 (1993).
\bibitem{24c}  A. Toukmaji, C. Sagui, J. Board, and T. Darden, {J. Chem. Phys.} \textbf{113}, 10913 (2000).
\bibitem{24d} M. P. Allen and  D. J. Tildesley, {Computer Simulation of Liquids} (Oxford University Press, New York, 1987).
\bibitem{24e} H. J. C. Berendsen, J. P. M. Postma, W. F. van Gunsteren, A. DiNola, and J. R. Haak, {J. Chem. Phys.} \textbf{81}, 3684 (1984).
\bibitem{24f}  M. Sprik, {J. Phys. Chem.} \textbf{95}, 2283 (1991).
\bibitem{29a} A. Kumar, {J. Phys. Soc. Jap.} \textbf{61},   4 (1992).
\bibitem{24g} K. Leung,   private commuincation (2011).
\bibitem{25}  G. Archontis and  E. Leontidis, {Chem. Phys. Lett.} \textbf{420},  199 (2006).
\bibitem{26}P.-A. Cazade,  J. Dweik, B. Coasne, F.  Henn,  and J. Palmeri,  {J. Phy. Chem. C} \textbf{114},  12245  (2010).
\bibitem{29} Ayse Arslanargin and Thomas L. Beck,  J. Chem. Phys. \textbf{136}, 104503 (2012).
\bibitem{30} Douwe Jan Bonthuis, Stephan Gekle, and Roland R. Netz, Langmuir \textbf{28},  7679 (2012).
\bibitem{31} Marcel D. Baer, Abraham C. Stern, Yan Levin, Douglas J. Tobias, and Christopher J. Mundy, J. Phys. Chem. Lett.  \textbf{3}, 1565 (2012).
\bibitem{32} Thomas L. Beck,  {Chemical Physics Letters} \textbf{561-562},  1 (2013).
\bibitem{30a}  {     D. Horinek, S. Mamatkulov, and R. Netz, J. Chem. Phys. \textbf{130}, 124507  (2009).}

\end{thebibliography}
\end{document}